\documentclass{aa}   

\usepackage{graphicx}
\usepackage{txfonts}

\usepackage{hyperref}
\hypersetup{
    colorlinks=true,
    linkcolor=blue,
    filecolor=blue,      
    urlcolor=blue,
    citecolor=blue
    }
\usepackage{natbib}
\usepackage[normalem]{ulem} 
\usepackage{xcolor} 

\begin{document} 

\title{Formation of an observed eruptive flux rope above the torus instability threshold through tether-cutting magnetic   reconnection}
\titlerunning{AR12241 flare}
\author{Avijeet Prasad\inst{1,2}
        \and
        Sanjay Kumar\inst{3}
        \and
        Alphonse C. Sterling \inst{4}
        \and 
        Ronald L. Moore\inst{4,5}
        \and
        Guillaume Aulanier\inst{6,1}
        \and
        R. Bhattacharyya \inst{7}
        \and 
        Qiang Hu \inst{5,8}
          }
   \institute{
   Rosseland Centre for Solar Physics, University of Oslo, Postboks 1029 Blindern, 0315 Oslo, Norway \\
   \email{avijeet.prasad@astro.uio.no}
    \and 
   Institute of Theoretical Astrophysics, University of Oslo, Postboks 1029 Blindern, 0315 Oslo, Norway
    \and 
    Department of Physics, Patna University, Patna 800005, India
    \and 
    NASA Marshall Space Flight Center, Huntsville, AL 35812, USA
    \and 
   Center for Space Plasma \& Aeronomic Research, The University of Alabama in Huntsville, Huntsville, Alabama 35899, USA
    \and 
    Sorbonne Universit\'e, Observatoire de Paris - PSL, \'Ecole Polytechnique, Institut Polytechnique de Paris, CNRS, Laboratoire de physique des plasmas (LPP), 4 place Jussieu, F-75005 Paris, France
    \and
    Udaipur Solar Observatory, Physical Research Laboratory, Dewali, Bari Road, Udaipur 313001, India
    \and 
    Department of Space Science, The University of Alabama in Huntsville, Huntsville, AL 35899, USA
    }

 
  \abstract
   {Erupting magnetic flux ropes (MFRs) are believed to play a crucial role in producing solar flares. However, the formation of erupting MFRs in complex coronal magnetic configurations and the role of their subsequent evolution in the flaring events are not fully understood.}
   {We perform a magnetohydrodynamic (MHD) simulation of active region NOAA 12241 to understand the formation of a rising magnetic flux rope during the onset of an M6.9 flare on 2014 December 18  around 21:41 UT (SOL2014-12- 18T21:41M6.9), which was followed by the appearance of parallel flare ribbons. }
   {The MHD simulation was initialised with an extrapolated non-force-free magnetic field generated from the photospheric vector magnetogram of the active region taken a few minutes before the flare.}
   {The initial magnetic field topology displays a pre-existing sheared arcade enveloping the polarity inversion line. The simulated dynamics exhibit the movement of the oppositely directed legs of the sheared arcade field lines towards each other due to the converging Lorentz force, resulting in the onset of tether-cutting magnetic reconnection that produces an underlying flare arcade and flare ribbons. Concurrently, a magnetic flux rope above the flare arcade develops inside the sheared arcade and shows a rising motion. The flux rope is found to be formed in a torus-unstable region, thereby explaining its eruptive nature. Interestingly, the location and rise of the rope are in good agreement with the corresponding observations seen in extreme-ultraviolet channels of the Atmospheric Imaging Assembly (AIA) of the Solar Dynamics Observatory (SDO). Furthermore, the foot points of the simulation's flare arcade match well with the location of the observed parallel ribbons of the flare. }
   {The presented simulation supports the development of the MFR by the tether-cutting magnetic reconnection inside the sheared coronal arcade during flare onset. The MFR is then found to extend along the polarity inversion line (PIL) through slip-running reconnection. 
   The MFR's eruptive nature is ascribed both to its formation in the torus-unstable region and also to the runaway tether-cutting reconnection.}


\maketitle

\section{Introduction} \label{sec:intro}
Eruptive events such as solar flares and coronal mass ejections (CMEs) are a manifestation of an abrupt and explosive release of energy stored in twisted magnetic field lines in the solar corona \citep{antiochos1999ApJ, Fleishman2020Sci}. 
In a magnetically dominated corona, these transient events generally  
relax the coronal magnetic field by releasing the stored magnetic energy {\citep{aschwanden2004book}}. Observations suggest that before the eruption, since the coronal magnetic field lines (MFLs) are rooted in the photosphere, they become stressed and twisted by the photospheric (rotational and shear) flows. Consequently, they end up storing the magnetic energy \citep{priest2002A&ARv, 2011LRSP....8....6S, priest2014book}. However, the physical process triggering the release of the stored energy still needs to be fully understood \citep{2011LRSP....8....6S, priest2014book}.

The magnetic structure central to the eruptive events is thought to be a magnetic flux rope (MFR), which is a bundle of twisted magnetic field lines winding about a common axis \citep{chen2011LRSP, priest2014book, he2022ApJ}.  Under favourable conditions, the eruptions are believed to be initiated by the sudden rise of the flux ropes. In general, the rise can be attributed to either ideal magnetohydrodynamic (MHD) instabilities of the MFR, such as the kink or torus instability {\citep{fan2007ApJ, kliem&torok2006prl, Demoulin2010ApJ}}, or it may be due to magnetic reconnection (MR); this is a process that changes the magnetic topology of MFLs and converts magnetic energy into heat, kinetic energy, and acceleration of the charged particles below the flux rope \citep{amari+2003apj, aulanier2012A&A, kumar+2016apj}.  
Therefore, it becomes imperative to understand the creation of the MFRs to understand solar eruptions. Typically, there are two standard scenarios proposed for MFR formation. In the first scenario, the flux rope is considered to be pre-existing below the photosphere and emerges into the corona during the flux emergence process {\citep{chen2000ApJ, fan_2001, fan_2010}}. In contrast, the second scenario suggests that the flux rope can be generated in the solar atmosphere through reconnections inside a sheared coronal arcade {\citep{1989ApJ...343..971V, 2001ApJ...552..833M, amari+2003apj, aulanier+2010apj, Jiang_2021Nat}}. These reconnection-based studies often rely on force-free fields and utilise prescribed flows at the bottom boundary to initiate MR. This paper also focuses on the process of flux rope formation through MR in a coronal arcade by simulating the dynamical evolution of a complex active region. However, in contrast to many previous works, here, the evolution is triggered by the Lorentz force (instead of the boundary flow) at the lower heights of the solar atmosphere. 

The instability of an MFR plays an important role in the eruptive nature of flares. For explaining eruptions in bipolar coronal arcades, the ``tether-cutting" model {\citep{2001ApJ...552..833M}
attributed the formation and subsequent eruption of the flux rope to runaway tether-cutting reconnection that begins when shearing and converging photospheric flows have brought oppositely directed field lines into close-enough proximity.
In this situation, ideal magnetohydrodynamic instabilities such as the ``kink instability" {\citep{torok2004, torok2005ApJ}} and the ``torus instability" \citep{kliem&torok2006prl, aulanier+2010apj} are also proposed to be causing the eruption. Recently, the modelling of eruptions has been further advanced using fully 3D MHD simulations {\citep{aulanier2012A&A, Janvier_2013, janvier2014, janvier2015SoPh}}. Their model shows the eruption of a flux rope through torus instability and reconnection at the current sheet that develops under the flux rope. Relevant to this, ``slip-running reconnection" has also been identified as a possible mechanism to trigger eruptive events {\citep{aulanier2006SoPh, Janvier_2013}}.  }

To construct an appropriate coronal magnetic field, the non-force-free-fields (NFFFs) extrapolation technique  {\citep{hu&dasgupta2008soph, hu+2008apj, hu+2010jastp}} is utilised. As the name suggests, the extrapolated field supports a non-zero Lorentz force, which is crucial for the spontaneous onset of the eruption. Consequently, the NFFF extrapolations differ from the widely used nonlinear-force-free-fields (NLFFFs) extrapolations which correspond to the vanishing Lorentz force equilibrium state \citep[e.g.][]{wiegelmann2008jgra,wiegelmann&sakurai2012lrsp,duan+2017apj}.
Utilising the NFFF extrapolations, recent MHD simulations successfully replicated the required coronal dynamics that lead to solar flares, coronal jets, and coronal dimmings {\citep{prasad+2018apj, nayak+2019apj, prasad_2020, Kumar-2022}}.  
In exploring the flux rope formation, we chose to numerically simulate the onset of the eruption of NOAA AR 12241, as the corresponding multi-wavelength observations suggest the origin of a flux rope over the polarity inversion line (PIL) \citep{joshi_2017}.
In this paper, we perform an MHD simulation initiated with NFFF extrapolation of AR 12241 to study the magnetic flux rope formation and the progress of its eruption leading to the flare. 
Interestingly, we find that the rope is formed during the time of pre-flaring activity of the M6.9 flare on 2014 December 18, and the subsequent rise of the flux rope coincides with the development of the parallel ribbons of the flare \citep{joshi_2017}. We find that our simulation results attribute the flux rope formation to tether-cutting reconnection that occurs at a low-lying hyperbolic flux tube (HFT) by the suitable converging Lorentz force. Moreover, a slip-running reconnection {\citep{aulanier+2005aa, aulanier+2010apj}} contributes to the extension of the flux rope along the PIL. 
The eruptive nature of the flux rope is attributed to two factors: the development of the flux rope in a torus unstable zone and the runaway tether-cutting  reconnection.

The rest of the paper is arranged as follows: Section~\ref{sec:observations} briefly mentions the important aspects of the flaring event, while Section~\ref{sec:nfff} provides details of the initial non-force-free extrapolated field. Section~\ref{sec:mhd} presents the simulation results and their relation to the multi-wavelength observations. The key findings of the paper are summarised in  Section~\ref{sec:summary}.

\section{Observations of M6.9 flare in NOAA AR 12241}\label{sec:observations}
\begin{figure*}[ht!]
\centering
\resizebox{\hsize}{!}{\includegraphics{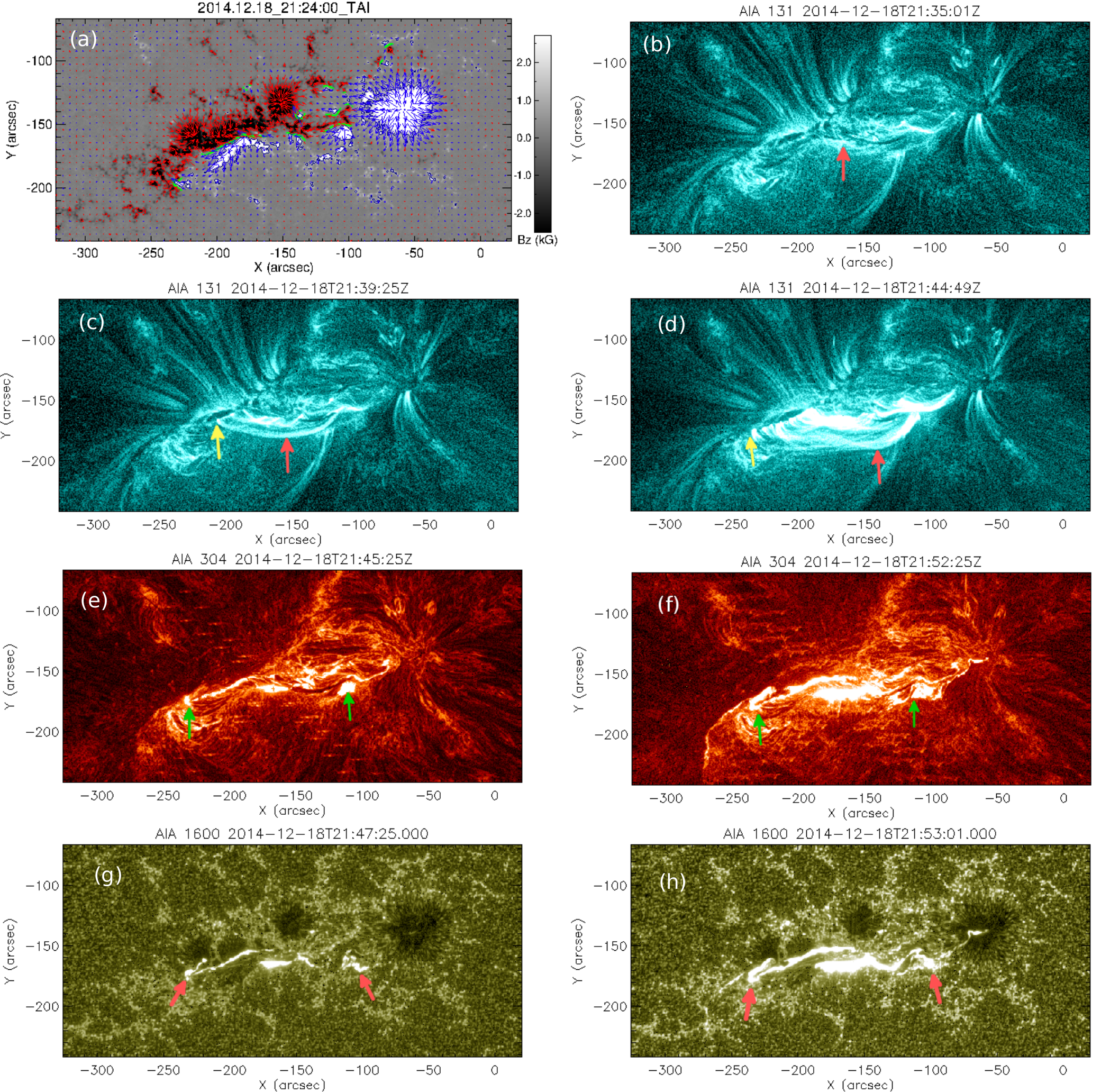}}
\caption{Depiction of the AR 12241 magnetogram, the formation and rising motion of the flux rope, and the parallel flare ribbons. Panel (a) illustrates the HMI vector magnetogram of AR 12241 at 21:24 UT on 2014 December 18. The grey-scale image is the vertical field, with the strength given by the colour bar. The red and blue arrows show the strength and direction of the transverse magnetic field, while the green lines mark the PILs. Panels (b)-(d) 
illustrate the flux rope formation, marked by the red arrow in panel (b) and its rising motion denoted by red arrows in panels (c) and (d) in 131~\AA. The brightenings moving eastward are marked by yellow arrows in panels (c) and (d). Panels (e) and (f) depict the parallel flare ribbons in 304~\AA. Also, note the isolated brightenings in 304~\AA~ (green arrows). Similar to 304~\AA, the parallel ribbons and the isolated brightenings are observed in 1600~\AA, as shown in panels (g) and (h), where the red arrows mark the isolated brightenings. North is upward, and east is to the left in this and all other solar images in this paper. 
The animations for panels (b) and (e) are available online.}
\label{f1-observations}
\end{figure*}

The M6.9 flare occurred in NOAA AR 12241 on 2014 December 18. A detailed study of the various observational aspects of the flare was already published by \citet{joshi_2017}, where these authors divided the flaring process into two components. The first component corresponds to the development of a flux rope over the PIL and its rise, leading to parallel flare ribbons. In the second component, the erupting flux rope reaches the three-dimensional (3D) magnetic null located at a much higher height and initiates MR at the null, forming the larger quasi-circular ribbon. This paper aims to simulate the dynamical evolution leading to the first component. For completeness, we first highlight the corresponding observations in Figure \ref{f1-observations}. The photospheric vector magnetogram of AR 12241 is from the Helioseismic Magnetic Imager \citep[HMI;][]{schou+2012soph} on board the Solar Dynamic Observatory \citep[SDO;][]{pesnell+2012soph} at 21:24~UT on 2014 December 18 (see Fig. \ref{f1-observations}(a)). The magnetogram is taken from the `hmi.sharp\_cea\_720s' data series \citep{bobra2014SoPh} that provides vector magnetograms of the Sun with a temporal cadence of 12 minutes and a spatial resolution of $0''.5$. 
To obtain the magnetic field on a Cartesian grid, the magnetogram is initially remapped onto a Lambert cylindrical equal-area (CEA) projection and then transformed into heliographic coordinates \citep{gary&hagyard1990soph}. We further cropped the field of view to 680$\times$340 pixels to focus on the region of interest. The dark green line shows the PIL, along which the flux rope is observed to develop.  

The flare starts at around 21:41 UT and peaks at $\approx$ 21:58 UT.
Figures \ref{f1-observations}(b)–(d) illustrate SDO/AIA 131~\AA~images. To adequately compare the simulation results with observations, SDO/AIA filtergrams are also CEA projected and remapped to the same spatial resolution as the magnetic field data, while the same field of view is used. Panel (b) shows the formation of the flux rope (marked by red arrow) at around 21:35 UT, and its subsequent rise and expansion can be identified in panels (c) and (d). Similar flux rope dynamics are plotted in Figure 5 of \citet{joshi_2017}.  Notably, there are the brightenings moving in the eastward direction (marked by yellow arrows in panels (c) and (d)). Panels (e) and (f) present SDO/AIA 304~\AA~images, which depict the creation of parallel ribbons during the flare. In addition to the parallel ribbons, the plots also document the development of isolated brightenings (marked by green arrows) on the eastward and westward sides of the parallel ribbons. SDO/AIA 1600~\AA~images further confirm the development of the parallel ribbons and the isolated brightenings (panels (g) and (h)). 
Also notably, at low heights, a cool-material filament also exists along the PIL {\citep{Wang_2017}}. However, its eruption during the flux rope formation and subsequent rise is not observationally established {\citep{joshi_2017}}. This suggests that its presence does not affect the flux rope dynamics. 

\section{Extrapolated coronal magnetic field of AR 12241}\label{sec:nfff}
\label{sec:3}

We utilise the non-force free extrapolation model developed by \citet{hu&dasgupta2008soph,hu+2008apj,hu+2010jastp} to extrapolate the coronal magnetic field of AR 12241 at 21:24 UT corresponding 
 to the vector magnetogram shown in Figure \ref{f1-observations}(a).
For the details of the model, readers are referred to \citet{hu+2010jastp} and references therein.
The model is based on the principle of minimum energy dissipation rate, according to which a plasma system prefers to relax toward a state that has minimum dissipation rate {\citep{bhattacharyya+2007soph}}. The magnetic field ${\bf{B}}$ is then shown to satisfy the double-curl Beltrami equation {\citep{bhattacharyya+2007soph}}. The field ${\bf{B}}$ then can be written as the superposition of three linear-force-free fields (one of which is selected to be potential in the model) \citep{hu&dasgupta2008soph}. Consequently, the Lorentz force associated with the field is non-zero. An iterative approach based on the minimisation of the average deviation between the observed and the calculated transverse field on the photospheric boundary is then employed to get the extrapolated NFFF.  
Recently, the NFFF model has been successfully utilised to explain various transient events in active regions such as flares, coronal jets, and coronal dimmings {\citep{prasad+2018apj, nayak+2019apj, prasad_2020}}.

Notably, the magnetogram displayed in Figure \ref{f1-observations}(a) has a spatial extent of $680\times 340$ pixels along
the $x$ and $y$ axes of a Cartesian coordinate system. To minimise the computational cost, the original domain is re-scaled to a
$320\times 160$ pixel grid in $x$ and $y$ directions. The vertical extension of the domain is taken to be $160$ pixels. In such a re-scaling, the inherent magnetic structures remain preserved {\citep{prasad_2020}}. In physical units, the size of the computational domain is roughly $245$ Mm $\times$ $122.5$ Mm $\times$ $122.5$ Mm along $x$, $y,$ and $z$.

\begin{figure*}[ht!]
\centering
\resizebox{0.95\hsize}{!}{\includegraphics{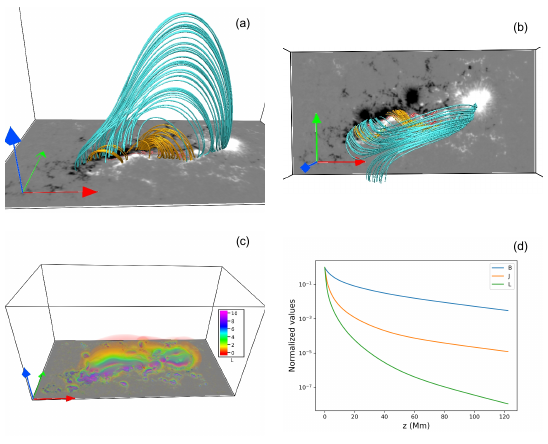}}
\caption{Extrapolated non-force-free field of the active region at the start of the MHD simulation of the onset of the flare-making eruption. Panels (a) and (b) depict the side and top-view of the extrapolated MFLs with the magnetogram as the bottom boundary. The low-lying yellow MFLs represent the sheared arcade over the PIL. Panel (c) shows the distribution of the relative magnitude of the Lorentz force density in the computational domain, documenting the existence of the force at lower heights near the PIL. The Lorentz force density has the units of $1.5 \times 10^{-8}$ dyne cm$^{-3}$.  Panel (d) illustrates the logarithmic variation of magnitude for the horizontally averaged magnetic field (B), the current density (J), and the Lorentz force density (L) with height, $z$. All of the values are normalised with respect to their maximum values.  }
\label{f5:initial_field_overview}
\end{figure*}

The magnetic field lines of the extrapolated initial field are plotted in Figure \ref{f5:initial_field_overview}. In this and subsequent figure, the $x$, $y$, and $z$ axes of the computational domain are represented by red, green, and blue axes, respectively. The red line in the $z=0$ plane denotes the PIL. The yellow magnetic loops mark the magnetic arches of the arcade over the PIL, with the side and top views shown in panels (a) and (b). Large-scale magnetic loops are plotted in cyan-coloured MFLs. Notably, there is the bifurcation in the field lines towards the southeast direction, more visible in the top view in panel (b), showing a sharp change in the field line connectivity. Noticeably, the cyan-coloured MFLs leave the computational domain through the lateral boundary ($y=0$). We performed an additional  extrapolation with an extended computational box along the lateral boundary (not shown) and found that these MFLs connect to the appropriate opposite polarity regions, validating the accuracy of the extrapolated field.
Panel (c) of Figure \ref{f5:initial_field_overview} shows the direct volume rendering of the Lorentz force density, which reveals the existence of the force at the lower heights and its sharp decay with height. Moreover, the panel also identifies the presence of a strong Lorentz
force in the vicinity of the PIL. Notably, the Lorentz force plays a crucial role in triggering the arcade dynamics, eventually leading to the MFR formation and the onset of the flare. Figure \ref{f5:initial_field_overview}(d) plots the variations of
horizontally averaged strength for the magnetic field, current density, and Lorentz force density with pixel height, $z$. 
The plots further confirm a faster decay of the Lorentz force density compared to the current density and the field strength. Consequently, the extrapolated field can be considered reasonably force-free at coronal heights, favouring the typical description of the solar corona.

\section{MHD simulation of AR 12241}\label{sec:mhd}

\subsection{Governing MHD equations  } 
The evolution of AR 12241 is studied by describing the coronal plasma by magnetohydrodynamics. With a focus on exploring the changes in field line topology, here we consider the plasma to be of uniform density, incompressible, thermally homogeneous, and perfectly electrically conducting \citep{kumar+2014phpl,kumar+2015phpl,kumar2017}. The set of non-dimensional MHD equations is then given as: 

\begin{subequations}
\begin{align}
\label{stokes}
&  \frac{\partial{\bf{v}}}{\partial t} 
+ \left({\bf{v}}\cdot\nabla \right) {\bf{ v}} =-\nabla p
+\left(\nabla\times{\bf{B}}\right) \times{\bf{B}}+\frac{\tau_a}{\tau_\nu}\nabla^2{\bf{v}},\\  
\label{incompress1}
&  \nabla\cdot{\bf{v}}=0, \\
\label{induction}
&  \frac{\partial{\bf{B}}}{\partial t}=\nabla\times({\bf{v}}\times{\bf{B}}), \\
\label{solenoid}
 &\nabla\cdot{\bf{B}}=0.  
\end{align}  
\label{e:mhd}
\end{subequations}

\noindent The magnetic field strength ${\bf{B}}$ and the plasma velocity ${\bf{v}}$ are normalised by the average magnetic field strength ($B_0$) and  the Alfv\'{e}n speed 
($v_a \equiv B_0/\sqrt{4\pi\rho_0}$ with $\rho_0$ representing the constant mass density), respectively. The plasma pressure, $p$, the spatial-scale, $L$, and the temporal scale, $t,$ are normalised by ${\rho {v_a}^2}$, the length-scale of the vector magnetogram, ($L_0$), and the Alfv\'{e}nic transit time, ($\tau_a=L_0/v_a$), respectively. 
Here, $\tau_\nu$ represents the viscous diffusion time scale 
($\tau_\nu= L_0^2/\nu$), with $\nu$ being the kinematic viscosity. 

The MHD equations are solved using the well-established numerical model EULAG-MHD \citep{smolarkiewicz&charbonneau2013jcoph}. The model details are described in \citet{smolarkiewicz&charbonneau2013jcoph} and references therein. Importantly, in the absence of the physical magnetic diffusivity (\ref{induction}), the dissipative property of the model intermittently and adaptively regularises the under-resolved scales by simulating magnetic reconnections. In our previous works \citep{prasad+2018apj, nayak+2019apj, prasad_2020, 2021PhPl...28b4502N, Kumar-2022}, we successfully simulated the dynamics of various active regions with the model and explained the solar transients such as flares, coronal jets, and circular brightenings in the active regions.  

\subsection{Computational setup}
\label{s:results}
The presented simulation was conducted in a computational domain having a $320\times 160 \times 160$ grid points for a physical domain spanning $[0,1]\times [0,0.5]\times [0,0.5]$ units in $x$, $y$, and $z$, respectively, where a unit length approximately corresponds to $245$ Mm. 
The simulation is initiated with the NFFF extrapolated magnetic field as shown in Figure \ref{f5:initial_field_overview}(a). With an initial zero velocity field, the simulated dynamical evolution is generated
 by the initial non-zero Lorentz force of the NFFF.  
 The resulting flow is incompressible, an assumption also utilised by \citet{dahlburg+1991apj} and \citet{aulanier+2005aa}. As our focus is to understand the flux rope's formation and early rising motion, this assumption seems justifiable in the tenuous coronal medium. 
 As there is no significant flux-emergence occurring during the event at the bottom boundary, we keep the $z$-components of the magnetic field ($\mathbf{B}$) and velocity field ($\mathbf{v}$) fixed to their initial values throughout the simulation. Magnetic reconnection is initiated at a certain height above the boundary, resulting in the plasma at the bottom boundary remaining perfectly ideal.
Under the constraint of incompressibility, which ensures no mass enters or leaves the domain at the boundary, these conditions effectively mimic the line-tied effects at the bottom boundary \citep{Jiang_2021Nat}.
For the other boundaries, we specified all the variables by linearly extrapolating their values from the interior points in their spatial neighbourhood \citep{prasad+2017apj, prasad+2018apj, 2020ApJ...892...44N}. This method of setting boundary conditions allows us to smoothly extend the conditions inside the simulation to the boundaries.

The value of the non-dimensional constant $\tau_a / \tau_\nu$ is chosen to be $2 \times 10^{-4}$, which is around 15 times larger than its coronal value. Higher values of  $\tau_a / \tau_\nu$ for the simulation only speed up the dynamical evolution, thus reducing the computational cost without any effect on the magnetic topology. The spatial unit step $\Delta x = 0.00625$ and time step (normalised by the Alfv\'{e}n  transit time $\tau_a \sim 30s$) $\Delta t = 2\times10^{-3}$ are selected to satisfy the Courant-Friedrichs-Lewy (CFL) stability condition \citep{courant1967jrd}.
The simulation is carried out for 6000 $\Delta t$, which roughly corresponds to an observation time of 45 Minutes. To better compare the simulated dynamics with the observations, we present the time ($t$) in units of $2 \tau_a = 1$ minute in describing the simulation results. 

\subsection{Evolution of AR12241}
\begin{figure*}[ht!]
\centering
\resizebox{0.95\hsize}{!}{\includegraphics{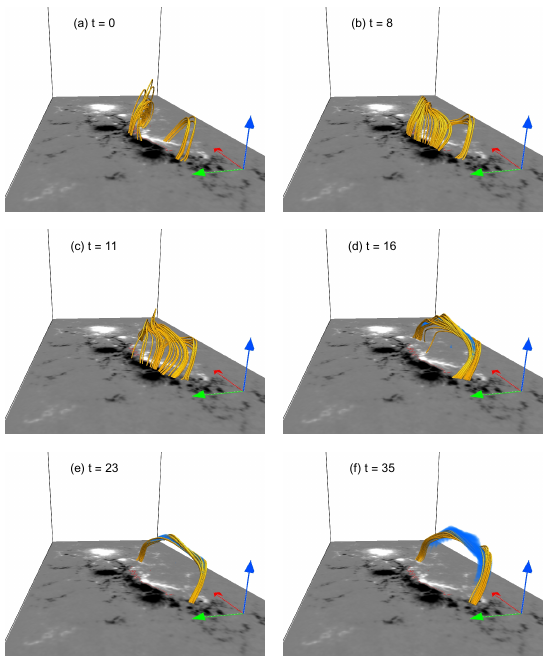}}
\caption{MHD simulation's evolution of the MFLs of the sheared arcade (in yellow). The bottom boundary is the magnetogram plotted in Figure 1. Important is the formation of the flux rope (panels (d)-(f)), which is evident from the appearance of a twist value that is greater than 1 (in blue).   As described in the text, $t$ has units of 1 minute, with $t$ = 0 corresponding to 21:24 UT. An animation of this figure is available.}
\label{flux-rope}
\end{figure*}

To understand the dynamics of the pre-flare stage of this event, we first describe the simulation's evolution of magnetic field lines of the arcade (in colour yellow) located in the vicinity of the event. The evolution is plotted in Figure \ref{flux-rope}. For plotting the evolution of the field lines, we employ the built-in ``field line advection"
technique of the VAPOR visualisation package {\citep{2005SPIE.5669..284C, 2008NJPh...10l5007M}}. In the technique, one representative point for a selected field line is advected by the velocity field and then the advected point is used
as a seed to plot the field line at a later time {\citep{2008NJPh...10l5007M}}. For a detailed description of the technique and its successful illustration in ideal as well as non-ideal magnetofluids, we refer to
\citet{2007NJPh....9..301C, 2008NJPh...10l5007M, li2019Atmos}.

\begin{figure*}[ht!]
\centering
\resizebox{0.8\hsize}{!}{\includegraphics{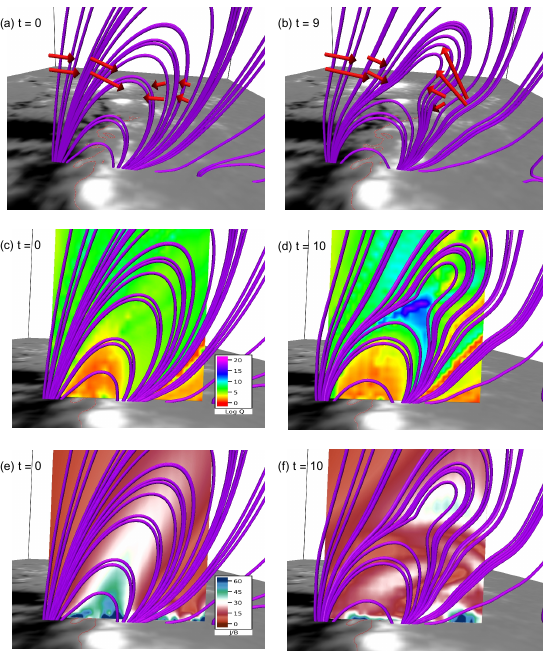}}
\caption{Simulation’s time evolution of the field-line map in a $y-z$ plane through the middle of the sheared arcade during eruption onset. The $z=0$ surface is the magnetogram of Figure 1(a). Panels (a) and (b) show the direction of the Lorentz force (red arrows), which is favourable for driving the anti-parallel field lines into close proximity and developing an X-type geometry. Panels (c) and (d) are superimposed with log Q, confirming the enhancement of $Q$-values co-located with the X-type geometry. Panels (e) and (f) are overplotted with $|\mathbf{J}|/|\mathbf{B}|$, which also shows an overall increase in the $|\mathbf{J}|/|\mathbf{B}|$ in the vicinity of the X-type geometry (panel (f)). 
   }
\label{fig4}
\end{figure*}

To explore the possibility of the flux rope formation,  the figure is further overlaid with the twist parameter (in colour blue) that measures the twist number of a field line. It is calculated by integrating the field-aligned current $\mathbf{J} \cdot \mathbf{B}/B^2$ along a field line \citep{torok2004, Berger_2006, liu+2016ApJ}. 
Figure \ref{flux-rope}(b) shows that the field lines of the arcade, rooted in the opposite polarity regions, are further sheared and stretched such that oppositely directed field lines approach each other.  This indicates that the initial Lorentz force is deforming the field lines of the arcade to give rise to the current sheet formation and consequent reconnection.
Subsequent evolution documents the generation of the twisted magnetic field lines at $t=11$. 
These field lines are co-located with the high values of the twist parameter, as shown in panels (d)-(f), suggesting the helical nature of the field lines. 
The magnitude of the twist parameter is around 1.05.
Consequently, the twisted field lines represent a magnetic flux rope generally located above the PIL {\citep{liu+2016ApJ, prasad_2020, Jiang_2021Nat}}.

\begin{figure*}[ht!]
\centering
\resizebox{\hsize}{!}{\includegraphics{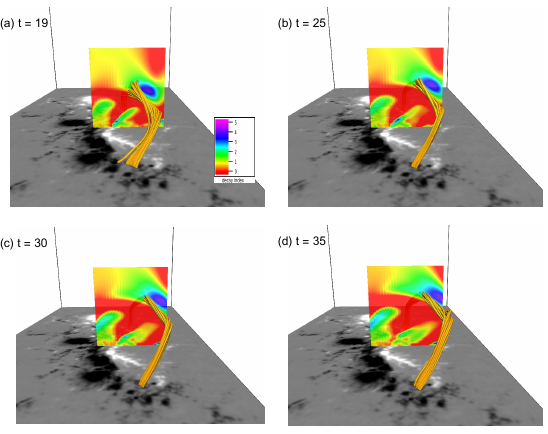}}
\caption{Dynamic rise of the flux rope with the decay index plotted in a $y-z$ plane through the middle of the flux rope. The z=0 surface is the magnetogram of Figure \ref{f1-observations} (a). The predominant part of the flux rope is co-located with a decay index greater than 2.}
\label{fig5}
\end{figure*}

To explore the physical mechanism underlying the flux rope formation and the role played in this by the orientation of the initial Lorentz force, in Figure \ref{fig4}, we plot the field lines of the transverse field (obtained by setting $B_x$=0 in ${\bf{B}}$). These field lines represent the magnetic field lines projected on the $x$-constant plane. The red arrows in panels (a) and (b) show the direction of Lorentz force at $t=0$ and $t=9$, respectively. 
Noteworthy are the foot-points of the field lines rooted on opposite polarity regions of the PIL. The Lorentz force is favourable to push the non-parallel field lines, located on opposite sides of the PIL, towards each other --- developing an X-type geometry in the projected field lines at a certain height (panels (b)). Such X-type geometry corresponds to a quasi-separator or HFT in the 3D magnetic field ($\bf{B}$), i.e. when the component $B_x$ is added in the transverse field {\citep{demoulin1996A&A, demoulin1997A&A, aulanier+2010apj,kumar_2021SoPh}}. To further confirm this, in panels (c) and (d) of the figure, we overlay the projected field lines with squashing factor ($Q$) \citep{Titov_2007,masson+2009apj,liu+2016ApJ} shown here on a logarithmic scale, which measure the field-line mapping of the magnetic field. The appearance of the high Q-values in the neighbourhood of the X-type geometry (Fig.~\ref{fig4}(d)) confirms the development of the HFT. 
Notably, the HFT with the high Q-values is often considered a potential location for facilitating magnetic reconnection \citep{titov2002JGRA,torok2004,masson+2009apj,aulanier+2010apj,kumar_2021SoPh}. This is also evident from panels (e) and (f) of the figure, in which the projected field lines are overplotted with $|\mathbf{J}|/|\mathbf{B}|$. 
As two non-parallel field lines approach each other closely in the vicinity of the reconnection site, there is an overall enhancement in $|\mathbf{J}|/|\mathbf{B}|,  $ seen in panel (f), suggesting a sharp increase in the magnetic field gradient.
Consequently, the scales become under-resolved, leading to magnetic reconnection in the simulation. These reconnections are responsible for the generation of the magnetic flux rope.
Further evolution documents the continuing reconnection at the X-type geometry that increases the magnetic flux inside the flux rope (not shown). It is worth mentioning that the post-reconnected field lines contribute to the formation of the flare ribbons.

\begin{figure*}[ht!]
\centering
\resizebox{\hsize}{!}{\includegraphics{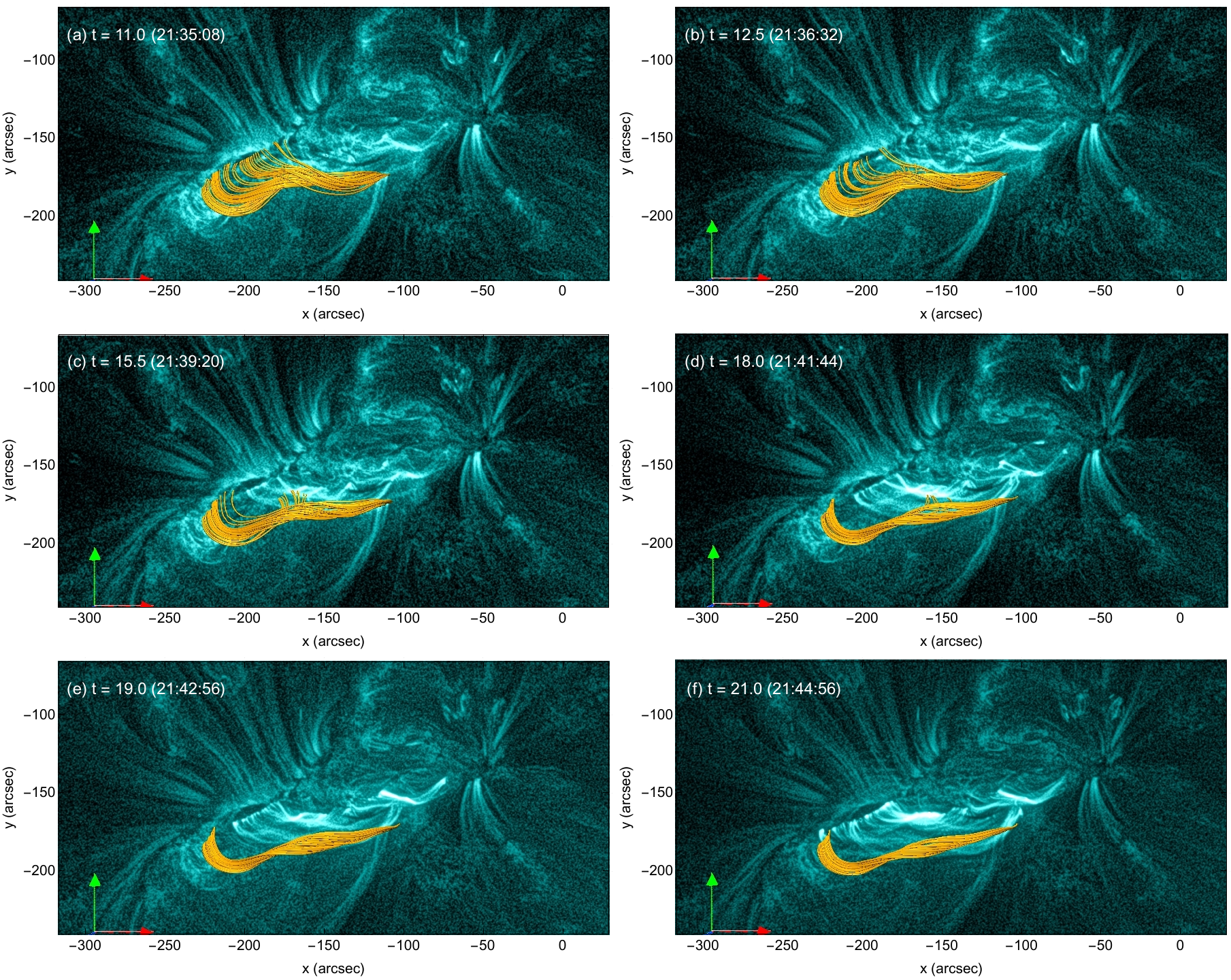}}
\caption{Superimposition of the simulated dynamic flux rope (in yellow) on co-temporal 131 ~\AA~images. The origin and rise of the simulated flux rope roughly match the observed ones.  The eastward movement of the eastern feet of the simulated MFLs of the flux rope and the observed brightenings are also apparent. The simulation time is given in the panels, while the corresponding time for the AIA images in UT is given in the parentheses. An animation of this figure is available.}
\label{131}
\end{figure*}

\begin{figure*}[ht!]
\centering
\resizebox{\hsize}{!}{\includegraphics{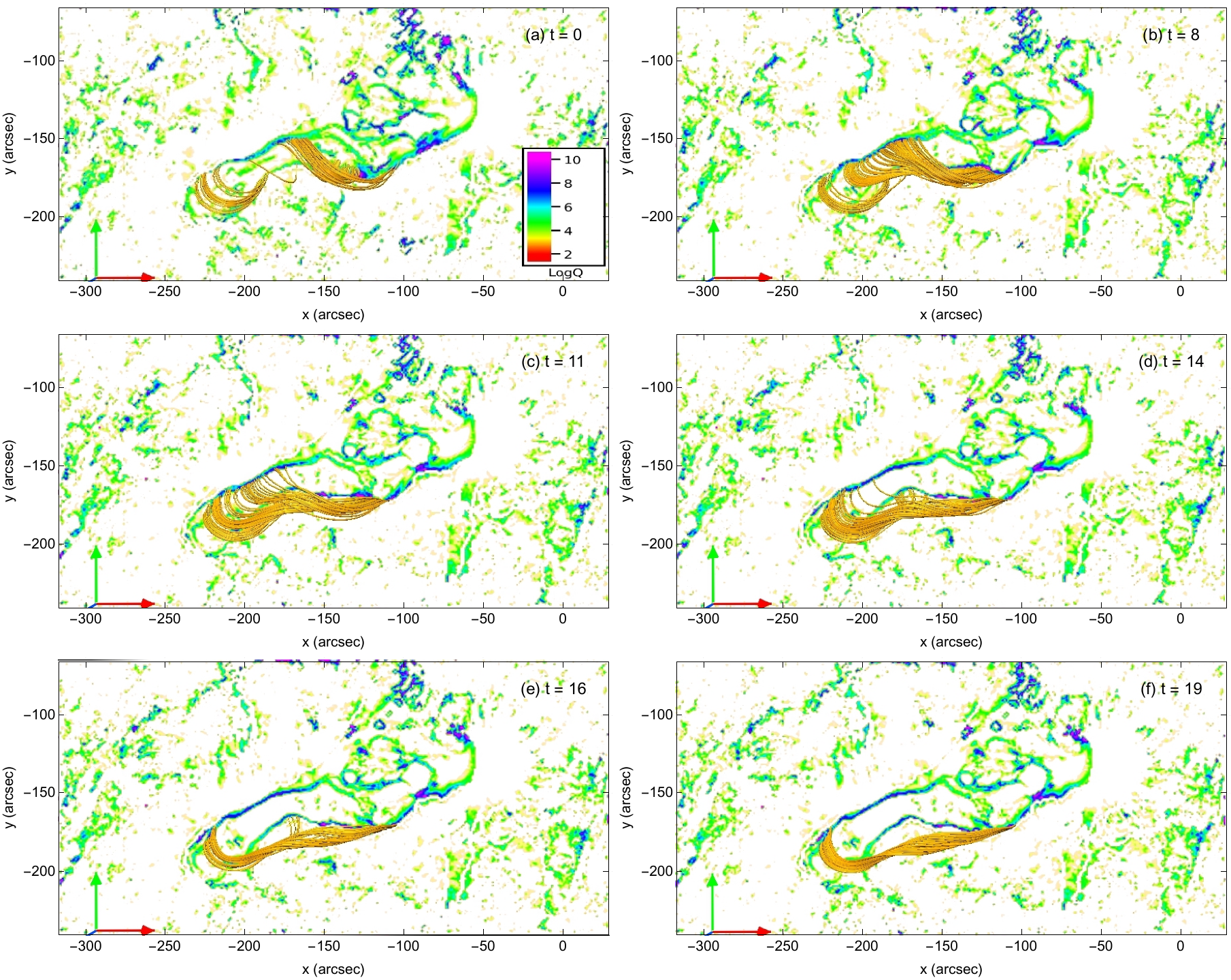}}
\caption{Dynamics of MFLs constituting the flux rope overlaid onto the $Q$-contours at the bottom boundary. Footpoints of the MFLs trace the large $Q$-contours. The motion can be attributed to slipping MR. An animation of this figure is available.}
\label{qsl-reco}
\end{figure*}

\begin{figure*}[ht!]
\centering
\resizebox{0.95\hsize}{!}{\includegraphics{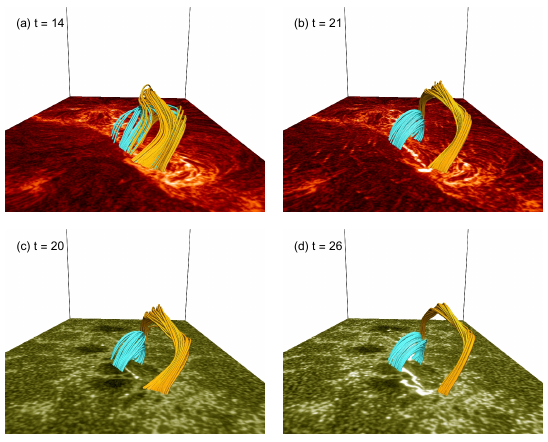}}
\caption{Evolution of the simulated MFLs of the flux rope and flare arcade, superimposed with the AIA 304~\AA, as shown in  panels (a) and (b), and~1600~\AA, shown in  panels (c) and (d), observations at the bottom boundary.  The appearance of the simulated post-reconnection arcades (in cyan) is evident, which are co-located with the observed parallel flare ribbons. The observed brightening near the foot of the simulated flux rope is also seen (in yellow). }
\label{304}
\end{figure*}

Relevantly, using the method proposed by \citet{fisher2012SoPh} for estimating changes in the integrated Lorentz force across the outer solar atmosphere by observing differences in the vector magnetograms, \citet{sarkar2019ApJ} computed the net change in the Lorentz force during a series of flaring events. Their findings, indicating the change in a range of $(2-5) \times 10^{22}$ dyne, pointed to a direct role of the Lorentz force at the onset of these events.
In our simulation, we determine the net change in the averaged Lorentz force values within a sub-volume of the flaring region, comparing the initial ($t=0$) and final ($t=35$) stages. The change in the averaged Lorentz force is $\approx 3.1 \times 10^{22}$ dyne, closely aligning with the estimates from observational studies. This supports the crucial role of the Lorentz force associated with NFFF extrapolations near flare initiation in triggering these events.
The MFR formation through MRs inside the sheared arcade appears to agree with the tether-cutting reconnection model \citep{2001ApJ...552..833M, toriumi2019LRSP}.
Moreover, the upward outflow generated by this reconnection carries the flux rope upward. The average speed at which the flux rope rises is around 10 km/s. This speed is much less than the typical coronal sound speed (200 km/s), which is the condition needed for the assumption of incompressibility to be valid.

To explore the eruptive nature of the formed flux rope, in Figure \ref{fig5}, in a $y-z$ plane passing through the flux rope, we depict the decay index in the vicinity of the rising flux rope. The decay index is a measure of the vertical decay of the strapping field overlying the flux rope.
The decay index is defined as $n= - d\log(B_t)/d\log(h)$, where $h$ is the height and $B_t$ is the transverse component of the overlying strapping field \citep{kliem&torok2006prl,Demoulin2010ApJ,liu+2016ApJ,jiang+2016nat}. Figure \ref{fig5} shows the rise of the flux rope from $t=19$ to $t=35$. It is noteworthy that the decay index is predominantly greater than 2 in the vicinity of the rising flux rope, suggesting the flux rope to be torus unstable \citep{kliem&torok2006prl, fan2007ApJ, Demoulin2010ApJ}. Consequently, in this case, the rising rope is eruptive. However, our simulation does not entirely capture the sudden rise of the erupting rope. This can be attributed to the assumed incompressibility condition. Moreover, the viscous relaxation also depletes the available free magnetic energy needed to generate the sudden rise.

To compare the simulated dynamics of the flux rope with observations, in Figure \ref{131}, the field lines of the flux rope are overlaid onto the observations in wavelength 131~\AA~at different times. Notably, the location and the time of the flux rope formation match with the observed ones, as shown in Figs. \ref{131}(a) and \ref{f1-observations}(b)). This suggests that the reconnections leading to the MFR formation are also responsible for the flare-onset activities. 
Also evident is the almost exact match of the simulated flux rope expansion and rise with the observations (cf. Figs. \ref{131}(b)-(f) and, Figs. \ref{f1-observations}(c)-(d)). Furthermore, Figure \ref{131} illustrates the motion of the eastern foot of the MFLs of the flux rope in the eastward direction, which appears to be in general agreement with the movement of the brightenings observed in the AIA 131~\AA~images, as marked by yellow arrows in Figs. \ref{f1-observations}(c)-(d).

To understand the movement, in Figure \ref{qsl-reco}, the evolution of the flux rope MFLs (plotted in Figure \ref{131}) is overlaid onto the $Q$-values at the bottom boundary. Notably, the motion of the MFLs is such that their footpoints always remain on the high $Q$-values ($log Q > 8$) contours. 
Such a motion can be attributed to slipping magnetic reconnections \citep{aulanier2006SoPh, joshi2017SoPh}. 
The brightenings in the 131~\AA~observations (Figure  \ref{131}) trace the high $Q$-contours, further corroborating the attribution. Notably, the apparent slippage motion is faster for the footpoints of MFLs rooted in negative polarity regions than those rooted in positive polarity regions (see the animation of Figure \ref{qsl-reco}). Such a difference in the motion can be ascribed to the different mapping norm \citep{Janvier_2013}.

Figure \ref{304} shows the eruption in AIA  304~\AA, shown in panels (a) and (b), and 1600~\AA, shown in  panels (c) and (d), images at different times, over-plotted with the simulated flux rope (in yellow). In the simulations, as the reconnection continues to build up the flux rope, the post-reconnection arcade also develops, marked by cyan-coloured MFLs in the figure. Notably, the footpoints of the simulated post-reconnection arcades are co-located with the observed parallel ribbons of the flare (see Figs. \ref{304}(b) and (d)). The accelerated charged particles produced at the reconnection site are then expected to travel along the arcade and decelerate in the denser chromospheric medium, leading to the formation of the ribbons {\citep{aschwanden2004book}}.
We also note presence of localised brightenings in both 304~\AA~ and 1600~\AA~observations (marked by cyan arrows in Fig.~\ref{f1-observations}(e)-(f) and red arrows in Fig.~\ref{f1-observations}(g)-(h)), which are near the foot of the eastern legs of the flux rope, as seen in panels (b) and (d) of Fig. \ref{304}. This indicates a causal connection between the brightenings and the flux rope formation.  

\section{Summary and discussion}\label{sec:summary}
This paper presents an MHD simulation of magnetic field evolution early in an M6.9 flare in AR 12241. The prime focus of the paper is to understand the formation of the magnetic flux rope during the initiation phase of the eruption that produced the flare. The initial magnetic field is generated by extrapolating the photospheric vector magnetogram of the active region obtained from HMI/SDO at 21:24 UT on 2014 December 18, using the non-force-free extrapolation technique. The extrapolated field has a Lorentz force at lower heights that becomes negligible in the corona, matching the standard picture of the coronal magnetic field. The Lorentz force, however, plays a crucial role in generating the self-consistent dynamical evolution from the initial static state. 

The initial non-force-free extrapolated field shows the presence of a sheared coronal arcade enveloping the PIL.
During the MHD evolution, under-resolved scales would evolve in the simulation as non-parallel MFLs come into close proximity with each other. The employed numerics then regularise these scales with simulated magnetic reconnections by producing locally adaptive residual dissipation. 
The simulated MFL evolution shows the origin of twisted field lines inside the initially sheared arcade at around 21:35 UT. The field lines represent a magnetic flux rope upon achieving a twist greater than one. 
The simulation demonstrates that magnetic reconnection at a low height in the solar atmosphere is responsible for the flux-rope formation and the M-class flare. The onset of that reconnection is attributed to developing favourable field line geometry in the form of a hyperbolic flux tube due to the suitable converging Lorentz force. Hence, the development of the flux rope is caused by tether-cutting reconnection at a low-lying HFT. In addition, the flux rope is formed in a region having a decay index $> 1.5$. This suggests that the newly formed flux rope is unstable via the torus instability and, therefore, has an eruptive nature. After its formation, the flux rope exhibits a slow rising motion while staying in the torus-unstable region.

SDO/AIA multi-wavelength observations (in particular,
the AIA 131~\AA~channel) of the M6.9 flare show the signature of the formation of a flux rope during the pre-flare stage. The initial formation location of the flux rope and its subsequent rise away from the photosphere matches the observations in 131~\AA, indicating that magnetic reconnection plays a vital role in the pre-flare activity. After its formation, the flux rope shows a rising motion and an extension with time of its east end.
In comparing our simulations with the SDO/AIA 304~\AA~and 1600~\AA~observations of the actual eruption, we find that our simulations closely fit the observed development of parallel flare ribbons and movement of brightenings with time towards the east. As noted above, the eastern feet of the MFLs of the flux rope show an eastward movement. This simulated movement coincides with the concurrent eastward expansion of brightenings in the 131~\AA~images. Interestingly, the simulation shows that the MFLs follow the high-value ln $Q$-contours; this suggests that slip-running reconnection is responsible for the motion and, hence, the brightenings. The generation of the post-reconnection arcade fits the parallel flare ribbons seen in the AIA 304 and 1600~\AA~images. Moreover, observable brightening also occurs near the eastern leg of the simulated flux rope, suggesting a causal connection between the flux rope and the brightenings. 

Overall, the presented simulation successfully captures the important observational features of the flaring event onset in the form of the flux rope development, the rise of the flux rope, and the early-flare arcade. This alignment of our simulation's results with observational data validates the initial assumption of our method, where we start with an unbalanced plasma state and allow the non-force-free field-induced Lorentz force to initiate the dynamical evolution of the active region. Moreover, the well-matched flare ribbons and post-flare loops, as depicted in our simulations, underscore the critical role of the Lorentz force in driving the reconnection processes leading to flare and eruption.
On the flip side, the current simulation was not able to demonstrate the fast-rise dynamics of the erupting flux rope, as the rising motion of the flux rope appears to cease towards the end of the simulation. This indicates that the runaway tether-cutting reconnection is not efficient enough by itself to lift the flux rope further. In the absence of prescribed photospheric boundary flows, the inefficiency of the reconnection can be attributed to the constantly depleting magnetic energy through the magnetic reconnection and the continuously operating viscous dissipation. 
Moreover, the incompressibility condition assumed in the simulation may also play a role in restricting the faster rise of the flux rope.  Relevantly, compressible MHD simulations driven by photospheric shearing flows or flux emergence have ascribed the eruption of flux ropes to the torus instability instead of the run-away tether-cutting reconnection {\citep{aulanier+2010apj, fan_2010}}. Therefore, the present simulation can be advanced further to simulate the eruption more realistically by relaxing the incompressibility and including the appropriate boundary flows, which are kept as future projects. 

\begin{acknowledgements}
We acknowledge using the visualisation software VAPOR (www.vapor.ucar.edu) for generating relevant graphics. Data and images are courtesy of NASA/SDO and the HMI and AIA science teams. SDO/HMI is a joint effort of many teams and individuals to whom we are greatly indebted for providing the data. AP would like to acknowledge the support from the Research Council of Norway through its Centres of Excellence scheme, project number 262622, and Synergy Grant number 810218 459 (ERC-2018-SyG) of the European Research Council. AP also acknowledges partial support from NSF grant AGS-2020703. QH acknowledges NASA grants 80NSSC21K0003, 80NSSC21K1671, and NSF grant AGS - 1954503. GA acknowledges financial support from the French national space agency (CNES) and from the Programme National Soleil Terre (PNST) of the CNRS/INSU, also co-funded by CNES and CEA. ACS and RLM were supported with funding from the Heliophysics Division of NASA’s Science Mission Directorate through the Heliophysics Supporting Research (HSR, grant No. 20-HSR20 2-0124) Program, and the Heliophysics System Observatory Connect (HSOC, grant No. 80NSSC20K1285) Program. 
We thank the referee for providing insightful suggestions, which led to the overall improvement of this paper.
\end{acknowledgements}

\bibliographystyle{aa} 
\bibliography{ms}

\end{document}